# Resistive hystersis effects in perovskite oxide-based heterostructure junctions

M.P. Singh

Laboratoire CRISMAT, CNRS UMR 6508, ENSICAEN,

6 Bd Maréchal Juin, 14050-Caen, France

L. Méchin

Laboratoire GREYC, ENSICAEN et Université de Caen, CNRS UMR 6072,

6 Bd Maréchal Juin, 14050-Caen, France

W. Prellier[1]

Laboratoire CRISMAT, CNRS UMR 6508, ENSICAEN,

6 Bd Maréchal Juin, 14050-Caen, France

M. Maglione

Institute of Condensed Matter Chemistry of Bordeaux, 33608-Pessac, France

**Abstract**

In this paper, we report the electrical and structural properties of the oxide-based metal/ferroelectric/metal (MFM) junctions. The heterostructures are composed of ultrathin layers of $La_{0.7}Ca_{0.3}MnO_3$ (LCMO) as a metallic layer and, $BaTiO_3$ (BTO) as a ferroelectric layer. Junction based devices, having the dimensions of 400 x 200 $\mu m^2$, have been fabricated upon LCMO/BTO/LCMO heterostructures by photolithography and Ar-ion milling technique. The DC current-voltage (I-V) characteristics of the MFM junctions were carried out. At 300 K, the devices showed the linear (I-V) characteristics, whereas at 77 K, (I-V) curves exhibited some reproducible switching behaviours with well-defined remnant currents. The resulting resistance modulation is very different from what was already reported in ultrathin ferroelectric layers displaying resistive switching. A model is presented to explain the datas.

---

[1] prellier@ensicaen.fr

Nowadays great deal of research activities are focused in the exploration of novel materials/properties for their utilization as memory and switching-based-devices. The current research activities have been driven by the ongoing miniaturization process of the silicon-based-devices and the demand for the increasing storage capacity per unit area for future electronic circuitries.[1] In principle, various physical characteristics (*e.g.,* the capacitance effects, resistance, magnetic spin dependent transports, and spin states) of the materials can be utilised for such a purpose.[2-10] Among the different possibilities, the hystersis effects observing in resistivity under external stimuli offer a great potential because of the simplicity in their read/write techniques. The material, which shows the hystersis effects in their resistivity, can be broadly classified in two categories. The first one corresponds to the materials in which the hystersis effect arises due to their electronic properties (i.e; they have internal effects). The second class of materials (doped or heterostructures) in which it is arising due to the external defects (*e.g.,* Cr-doped $SrZrO_3$ films on Si)[10] or interfacial effects (heterostructures/multilayers)[4] which have been designed by artificial means (i.e., they have external effects). As an example in devices fabricated on $Pr_{0.7}Ca_{0.3}MnO_3$ using the Ag-as electrode, a resistive hysteresis effect is observed under application of an electric pulse [2-3]. Such results have a potential utilization for switching/memory based devices.[2-7] They are basically the internal properties of a given materials. However, it is important to note that the electronic properties of the manganites (*e.g.,* charge and orbital ordering) are very much dependent on the films growth conditions.[11] On the other hand, the artificial heterostructure offers a great potential because of the quality of the structure and an easy control of the interface. Furthermore, their properties and devices characteristics can be tailored and optimized for suitable applications by a careful control of its structure. Since it is an external effect, the hysteresis effect can be tailored by optimizeing the junctions' dimension, its type and its nature. This has motivated our study to investigate the artificial nano-heterostructures.

In recent years, various groups have proposed the possibility of designing the heterostructures-based-devices by combining the manganites and ferroelectrics.[2-7] In these devices, the hystersis effects are observed at high electric field whereas the device, which we are going to present here, shows a hysteresis effects at low field. Consequently, it offers a greater potential. More precisely, our samples are composed of ultrathin layers of all oxides-based metal-ferroelectric-metal junctions in contrast to the previous studies where junctions have been formed using the base metal (*e.g.,* Au), ferroelectrics, and oxide-metal. In the present study, the artificial heterostructures have been first designed by employing the $La_{0.7}Ca_{0.3}MnO_3$ (LCMO) as a metallic electrode and $BaTiO_3$ (BTO) for the insulating layers. Various techniques have been employed to characterise the junctions and the results are presented in the article. At the end, an explanation on the origin of the present hysteresis is given, discussed and compared to previous reports.

Tri-layers of $La_{0.7}Ca_{0.3}MnO_3/BaTiO_3/La_{0.7}Ca_{0.3}MnO_3$ were grown on (001)-oriented $SrTiO_3$ (STO) substrates using the pulsed laser deposition technique. The growth was done at 720°C, in a flowing oxygen at 100 mTorr. Polycrystalline, stoichiometric targets of LCMO and BTO were used for ablation. These targets were synthesised by the standard solid state routs using their appropriate binary oxides. The 250 Å thick LCMO layer was first grown, followed by the growth of BTO layer (having a thickness in the range of 30-70 Å). A 250 Å thick LCMO layer was finally grown on top of BTO to end up the structure. To fabricate the devices, and minimize the contact resistance, a 500 Å thick layer of Au was also deposited on top of the tri-layer structure, at room temperature by RF sputtering technique, through shadow mask. A schematic drawing of the structure is shown Fig. 2a. Through photolithography techniques and Ar-ion etching, junctions (dimensions of 400 μm x 200 μm) were patterned. The etching process was stopped so that the bottom LCMO layer was not etched. The structural quality of the samples was made by X-ray diffraction (XRD) using Cu Kα radiation ($\lambda$=1.5406 Å). DC electrical properties of films were measured in four probe configuration using a HP 4156B precision semiconductor parameter analyser.

Bulk LCMO is a ferromagnetic metal with the Curie temperature of 250 K. It exhibits a pseudo-cubic structure with a lattice parameter of 3.86 Å. Likewise, BTO is a ferroelectric in nature with the Curie temperature of 400 K. It is cubic with the lattice parameter of 4.006 Å in its paraelectric states, whereas it crystallizes in a tetragonal structure in ferroelectric states with lattice parameters a=3.996 Å and c = 4.01 Å.[12, 13] Thus, from the lattice parameter point of view, the difference between the cubic and the tetragonal structure is not very large. The mismatch between the LCMO and BTO lattice parameter is about ~ 2 %, whereas it is about 1.7 % with respect to the $SrTiO_3$, which possesses cubic symmetry with the lattice parameter of 3.905 Å. This indicates that it is possible to grow the heterostructures of LCMO and BTO layers deposited on STO substrates. Since the structures are very pseudo-cubic, they should also be epitaxial when grown on top of each other.

To examine the crystal structure of the as-grown tri-layers, X-ray diffraction study was performed. Fig. 1a show a typical θ-2θ XRD pattern of the tri-layers composed of 70 Å of BTO recorded in the 10-100° range. The diffraction study shows several peaks corresponding to LCMO, BTO and STO substrate, indicating the film is single phase without any impurity. The full-width-at-half-maximum (FWHM) of the rocking curve, recorded around the fundamental (002) diffraction peak of the LCMO, is very close to the instrumental broadening (0.3°), confirming the good-crystallinity. The presence of weak BTO peaks, with an out-of plane lattice parameter close to 4.1 Å, was attributed to low thickness of the layer. To study further the coherence of the films, pole figure were recorded around the asymmetric (*10l*)-reflection of LCMO, where *l = 1, 2, 3, 4*. The pole figure of the film (Fig.1b) composed of 70 Å thick BTO displays are only four peaks, which demonstrates that the films have grown cube-on-cube. Thus, this structural study clearly demonstrates that the tri-layers has grown coherently, with well-defined interface.

Several junctions have been fabricated with different areas, and the transport in current-perpendicular to the plane (CPP) configuration was thusly measured. Fig 2b and 2c show the typical (I-V) characteristics recorded with current biased at 300 K and 77 K, respectively. The (I-V) characteristics at room temperature (Fig 2b) clearly show the linear behaviour and may be understood by the leakage current driven by the thermal process. The corresponding resistance is about 34 kΩ. With progressive decrease in the temperature below 200 K, there is an appearance of a hysteresis in (I-V) curves. This feature becomes very prominent below 100 K. As can be seen from the Fig.2c with increase the voltage, there is constant current flow. Suddenly, a sharp increase of the current is observed at ~ +1 V and approaches to a divergence. Reversing of the electric filed results in remnant currents and draws a loop, with a very sharp current reversal at voltage ~ -1V. These hystersis loops in I-V and consequently the hystersis in resistivity (Fig. 3) are very much traceable from time to time and thus, reproducible. After heating the sample to room temperature, and again cooling down to 77 K, the sample shows the identical I-V characteristics as exhibited in Fig 2c. Thus, the present (I-V) curves show well-defined hysteretic loops. In fact, the switching nature in the (I-V) curves have also been reported in some other oxides-based devices. In the case of Pt-PbZr$_{0.52}$Ti$_{0.48}$O$_3$-SrRuO$_3$, the hysteretic features have been attributed to the ferroelectric nature of the PbZr$_{0.52}$Ti$_{0.48}$O$_3$.[4] Nevertheless the major difference, between the present data and the reported ones, is the low field hysteresis and its symmetrical nature. This symmetric hysteresis loops was not observed with the Pt- PbZr$_{0.52}$Ti$_{0.48}$O$_3$-SrRuO$_3$ junctions. In these PbZr$_{0.52}$Ti$_{0.48}$O$_3$ films, the loops are symmetric but with a strong fading of the current close to the zero-bias and, two opposite loops at positive and negative voltage. Such behaviour is expected for a purely resistive switching (which requires zero current at zero bias). This is very different from our results where non-zero currents at zero bias are observed, on increasing or decreasing sweeps. Thus, these non-zeros remanent currents in our LCMO/BTO/LCMO trilayers suggest that the switching loops are not purely resistive. At low bias, the currents result from the displacement nature due to the polarisation switching within the ferroelectric layer. We understand the present characteristics as follows. In the case of a ferroelectric loop, the displacement currents require a capacitor in series to be integrated in the detection system. Built in blocking capacitors at the BTO/LCMO interfaces could act as such integrating capacitors at low voltages. When the bias is raised above a threshold (of about ±1V), these blocking layers switch to a conducting state leading to the observed current divergence. At the end, the stack switches from a ferroelectric behaviour at low voltages to a conducting state above the threshold bias. In this way, the interfaces act as diodes placed in opposite directions. The inner ferroelectric loops are not switched because the coercive field required to reverse the polarisation has not been reached, since the interface conducting state occurs before the screening of the effective internal field.

The next difference between the inner part of our ferroelectric loops lies with the dynamics of the observed currents (Fig 2c). At fast speeds, the remanent currents are stronger and a sharp decrease is observed on slowing down the measuring rate. This means that the polarisation inducing displacement-currents are not permanent. A strong and slow relaxation of the remanent currents also occur. In fact, this kinetics may be a direct consequence of the charge accumulation process at the interfaces since usually, in standard semiconductor diodes, such accumulation process is very rapid because electrons and holes act basically as accumulating species.[13] On the other hand, in the case of BTO/LCMO interfaces, it is governed by the various factors, e.g., defects such as oxygen vacancies or 2-dimensionnal defects or, intrinsic interaction between the electric dipoles and the free oxide charges. These elements lead to the slowing down of the accumulation process.[14-16] In ferroelectric thin films such as $PbZr_{0.52}Ti_{0.48}O_3$, one can see ferroelectric hysteresis loops shifting due to interface charge accumulation.[14] In single crystals and ceramics, the slow space charge processes have also be evidenced.[14-16] At the present time, we cannot rule out a purely space charge accumulation as a possible source of the ferroelectric-like hysteresis loops of Fig.2c. In this case, the non-zero remanent currents would result from transient space charge dipoles stemming from unbalanced space charges at the two BTO/LCMO interfaces. Impedance spectroscopy and dielectric experiments are underway to probe such space charge dipoles and their thermally excited relaxation. However, the other alternative possible explanation is not ruled out.

By lowering the temperature, the other important feature (Fig. 3) observed in these junctions is the modulation of resistance across junctions by several orders of magnitude. Usually multilayer composed LCMO/BTO superlattices exhibited the low resistivity (~$10^6$ Ω-cm) compared to the present tri-layer junctions (~$10^{10}$ Ω-cm). This can be understood based on the recent work on the field effect transistors utilizing $Nd_{0.7}Ca_{0.3}MnO_3$ as a channel and $PbTiO_3$ as gate dielectrics by T. Venkatesan and his co-workers[7]. They have clearly shown that the resistance of the manganites can be modulated due to the non-linear electric fields of ferroelectric layers, *i.e.,* $PbTiO_3$. Thus, the present hysteretic (I-V) characteristics of LCMO-BTO junction's open the route to understand the current through ferroelectric junctions and may help to design the new non-volatile memory devices based on the resistivity switching.

To summarize, we have successfully grown the tri-layers $La_{0.7}Ca_{0.3}MnO_3/BaTiO_3/La_{0.7}Ca_{0.3}MnO_3$ heterostructure on $SrTiO_3$ (001) by the pulsed laser deposition technique. The electrical characterization of the devices, based upon these trilayer structures, exhibit the hysteresis in (I-V) curves with well defined remanent currents at zero bias followed by a strong resistance-decrease at a threshold voltage. The observed switching states in (I-V) curves have been understood either as possible switching between the ferroelectric domains of $BaTiO_3$ or by a charge accumulation process at the interfaces.

This work has been carried out in the frame of the European STREP project "Manipulating the Coupling in Multiferroic Films" MaCoMuFi (033221) and the European Network of Excellence "Functionalized Advanced Materials Engineering of Hybrids and Ceramics" FAME (FP6-500159-1) supported by the European Community, and by Centre National de la Recherche Scientifique (CNRS). Partial support from the Agence Nationale de la Recherche (NT05-3_41793, NT05-1_45147) is also acknowledged.

**Figure captions:**

**Figure 1.** θ-2θ XRD patterns of the as-grown tri-layer heterostructure grown on SrTiO$_3$ substrates (STO). (b) Corresponding pole-figure recorded around the asymmetric (103)-LCMO reflection. Note the sharpness and intensity of the diffraction peaks attesting a good crystalline quality.

**Figure 2.** (a) A schematic diagram of the device structure after the photolithography and Ar-ion etching. Bi-directional DC (I-V) characteristics of the devices measured at (b) 300 K and (c) 77 K. The corresponding resistance is 34 kΩ. The (I-V) data were collected under various acquisition times.

**Figure 3.** Variation of device resistance at 77 K as a function of voltage derived from the respective I-V curves (Fig 2c). A modulation of resistance across junctions by several orders of magnitude is oberserved. Arrows are just for visual guide and it show the direction of I-V curves traced.

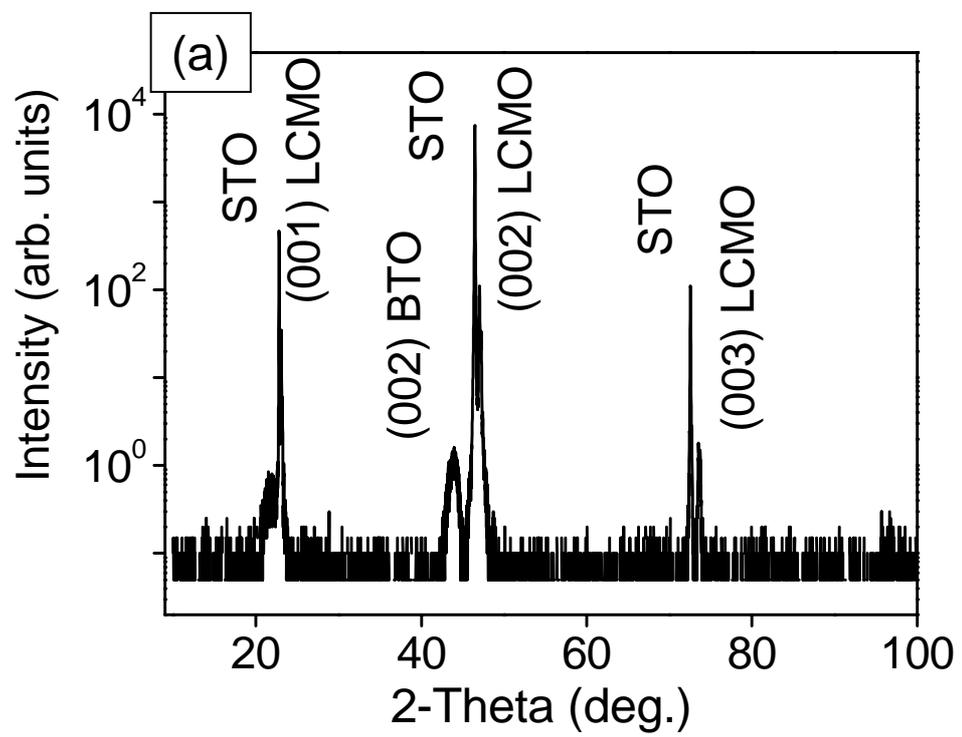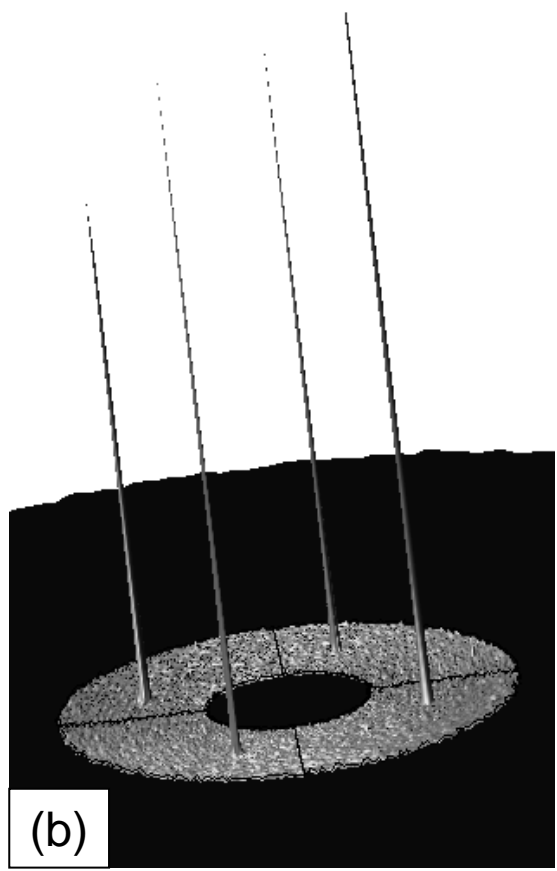

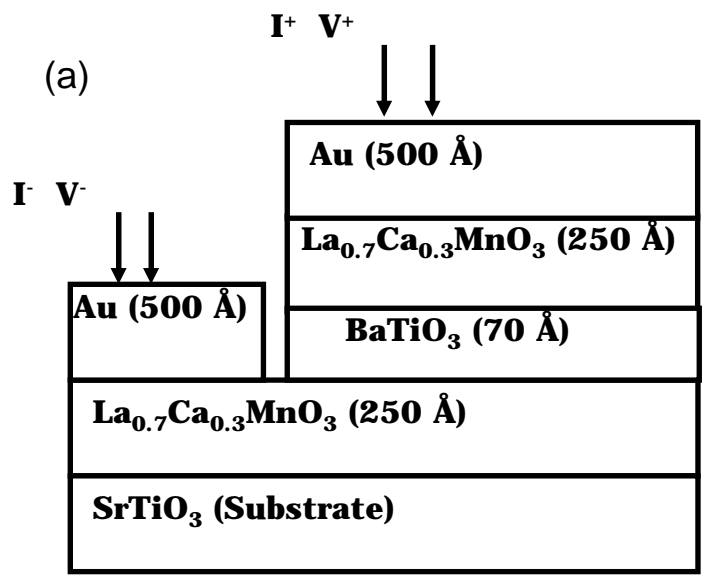
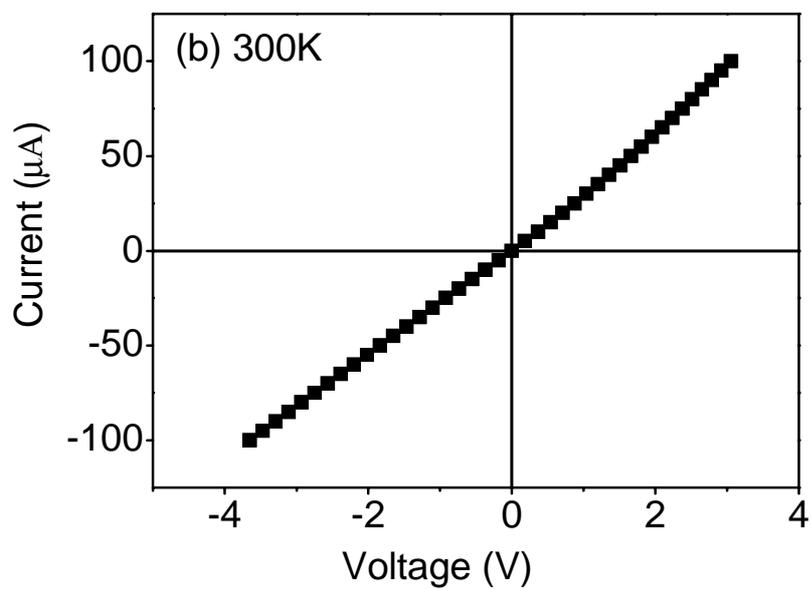
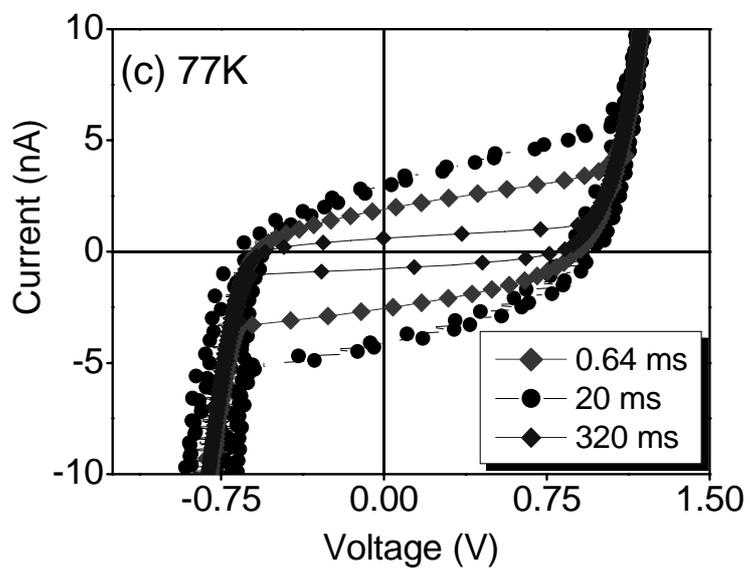

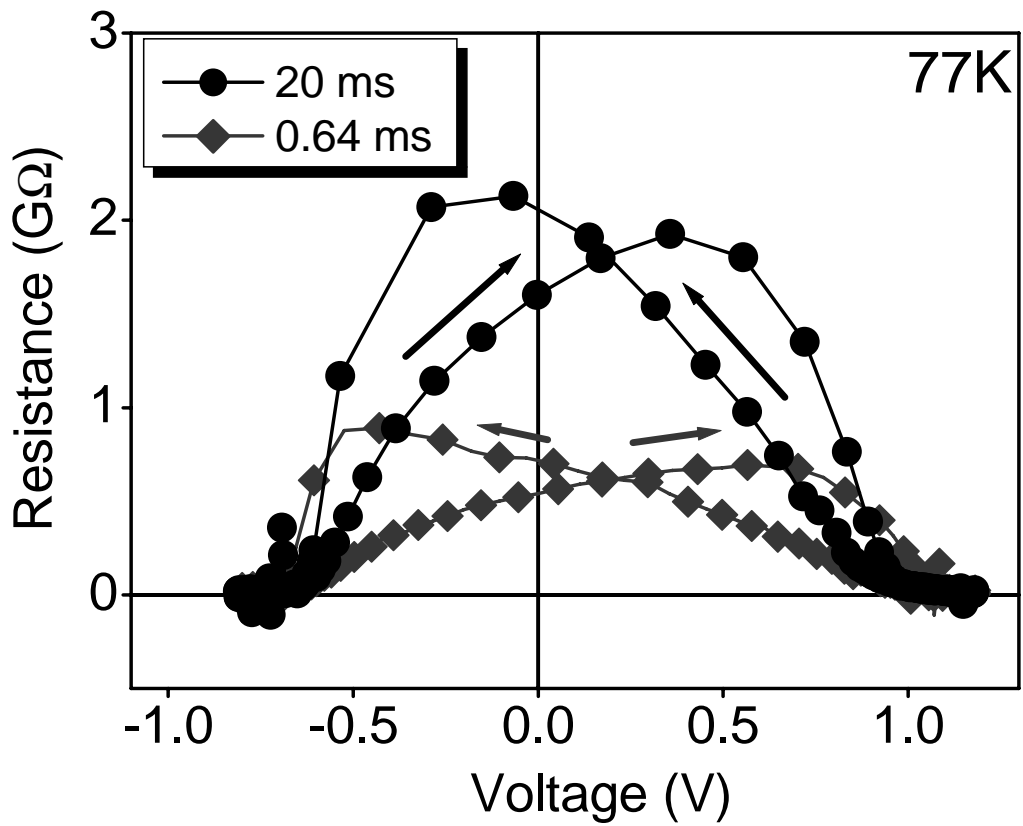